\documentstyle[12pt]{article}
\title{Gauge Theoretic Chaology. \footnote{ Published in "Non-Linear Dynamical Systems", eds: V.Srinivasan, A.K. Kapoor \& P.K. Panigrahi, Allied Publishers, 2000.}}
\author{ Bindu A. Bambah\\
School of Physics, University of Hyderabad, Hyderabad-500 046.\\
C.Mukku\\
CAS in Mathematics and Dept. of Computer Science\\
Panjab University, Chandigarh.\\
M.S.Sriram\\
Dept. of Theoretical Physics,
University of Madras, Chennai.\\
S. Lakshmibala\\
Dept.of Physics,
I.I.T, Chennai.}

\date{ 5th November 1999}

\begin{document}

\maketitle

\abstract{We review here the work done on  the occurence of
chaotic configurations in systems derived from Gauge theories.
These include Yang-Mills and associated field theories with
modifications including Chern Simons and Higgs fields.}

\section{Introduction.}

The upsurge of interest in studies of Chaos in various systems has all the ingredients
of a `scientific revolution' as defined by Thomas Kuhn in his pioneering work `The Structure of Scientific
Revolution'. Kuhn describes a scientific revolution as the result of a
dramatic shift of "paradigm". A paradigm is an "accepted example of actual scientific practice
from which spring coherent traditions of scientific research". Transformation of paradigms lead to scientific revolutions,
which in turn lead to cross fertilization between different fields,  giving rise to new areas of research.
The chaos revolution represents a shift from  calculus and a smooth deterministic
description of dynamical systems to  a highly non-linear, often fractal and  unpredictable one.
Thus, in some sense, chaos is a revolution from a deterministic world view to an unpredictable picture of the universe resulting from non-linearity.
Since non-linearity of physical and biological systems is more of a rule than an exception,
the possibility of applications is enormous.

This mathematical revolution was paralleled in physics by another revolution
with the advent of quantum field theory and the postulation of a
"fundamental theory of particles" based on the "paradigm" of gauge symmetry.
This field, the modern avatar of which is "string theory",  has lead to
new era in applied mathematics, in which not only have fundamental mathematics
been applied to solve physical problems, but physical concepts have
contributed to solving problems in pure mathematics and generating
new areas of mathematics.
Of particular relevance to this review,  are Yang-Mills theories and their topological
generalizations vis-a-vis Chern Simons terms and the profound impact on mathematics they have had
since the work of Atiya-Drinfield-Hitchin-Manin and that of Witten and Donaldson recently.
Yang-Mills theories, central to quantum theories of elementary particles, became of interest
to dynamicists with the recognition of the fact that the Self Dual Yang Mills theories
serve as some sort of a "universal integrable system" from which a large class
of known integrable systems can be derived by reduction.

The belief that chaotic phenomenon play
an important role in the study of fundamental particles
goes as far back as Heisenberg , before the advent of gauge theories,
in a seminal paper describing meson production in terms of turbulent fields.
As the modern theory of fundamental particles is based on
 Yang-Mills Field theories  which are highly non-linear and evolve chaotically in space-time,
work in the dynamical structure of these theories was initiated in the eighties by Matinyan , Saviddy and their collaborators $[6]$,$[9]$
in an attempt to study the role of chaotic phenomena in quark confinement.
Since then a whole spectrum of work has been done in chaos in Yang-Mills theories
and its various extensions, including topological ones,  and contributions of chaos
to particle production processes as well as confinement have been studied.
This review is intended to survey this field, highlighting the work done, its implications and
future studies that can be done.
Indeed, it represents and area in which the two hitherto parallel revolutions in Mathematics and Physics
cease  being parallel and converge.

In our recent papers $[1]$,$[2]$,$[3]$,$[4]$,$[5]$, we have continued these studies into chaos in Yang-Mills and associated field theories.
Of special interest in the modern context and in our work is the role
of  topological Chern-Simons  "effects" in dynamical systems theory.
The Chern Simons term is the metric independent "topological " term that
can be added to Yang Mills theories in odd-dimensions. It is a different way to get finite mass
other than the Higgs mechanism. This mass is hence called the topological mass.
We have shown , as will be described in the review , that a field theory described by a Lagrangian
of the type $L=L_{YM}+L_{CS}+L_{Higgs}$ admits order chaos transitions as a function of energy and the three
parameters (the gauge coupling, the "topological mass", and the "Higgs coupling) of the theory.
The dynamical systems are obtained from  the field equations of $L$ through an assumption of spatial
homogenity. Earlier workers in the field had placed additional restrictions and reduced the dynamics to two degrees of freedom.
In our work, we have maintained the full complement  9 degrees of freedom for the SU(2)
gauge group. The results therefore are mainly numerical out of necessity, though a thorough Painleve analysis,
which will be described has been done. We find from our numerical results that the order chaos transition in this system implies a constant creation and destruction of KAM torii
in phase space. Thus far this phenomenon has not been seen in Hamiltonian systems and warrants an analytical study.

In addition to physical applications the study of the Chern-Simons term has the additional benefits in dynamical systems study, in that, being topological
it admits topological invariants in its quantum analysis. Since this topological theory lies in three dimensions, it yields topological information on 3-manifolds,
in particular on knots and links embedded in three dimensions.
It has been conjectured that , viewed as a dynamical system, the resulting ODE's derived from
Chern-Simons field theory may be used to generate knots and links by seaching for periodic orbits and/or strabge attractors
in the sysem. An alternative, more automated method would be to generate time-series data
from the ODE and use simple recurrence to find periodic orbits and the Ruelle-Takens method to search for a strange attractors in the data.
The strange attractor would then allow a knot template to be constructed and the associated embedded knots to be enumerated and examined.
These are just some of the directions in mathematical physics that a study of
the dynamical behaviour of topological systems can lead us to.
The review presented here is a summary of the work already done in this field.

We begin this  review with an overview of the work done on Yang-Mills dynamical systems and  their extensions. We present the analytical and numerical tools
for the study of these dynamical systems and then sketch the work done which extends to the entire field theory by
considering perturbations around vortex solutions of the field theory.We conclude by highlighting
the salient results and highlighting the avenues for future research.

A useful table of identification of Gauge theory terminology with Fibre Bundle Terminology due to C.N. Yang
is given in Table 1.

\begin{table}
\centering
\caption{Translation of Terminology from Physics to Mathematics}
\vspace{0.5in}
\begin{tabular}{||l|l||}  \hline \hline

{\bf Physics.}  & {\bf Mathematics.}      \\ \hline \hline
Global gauge&Principal Co-ordinate Bundle\\ \hline
Gauge Type&Principal Fibre Bundle\\ \hline
Gauge Potential $A_{\mu}^{\lambda}$&Connection on a Principal Fibre Bundle\\ \hline
Field Strength $F_{\mu \nu}$& Curvature\\ \hline
Electromagnetism& Connection on a U(1) Bundle\\ \hline
SU(2) Yang Mills&Connection on an SU(2) bundle\\      \hline
Dirac Monopole& Classification of U(1) bundle according to first \\
&Chern Class.\\ \hline
\end{tabular}  \\
\end{table}

\section{ Motivation and History of Problem.}
In considering a quantum field theory of non-Abelian fields,
quantization of fields proceeds by examining fluctuations around a particular
classical solution and building up a Hilbert space on which the operators of
the quantum theory then act.
Quantum theory is probabilistic in nature and therefore, it is necessary that
the classical solution chosen for quantization be of lowest energy otherwise,
there would be a finite, non-zero probability of the quantum solution decaying
into the state of lower energy.
For Yang-Mills theories, in particular when the gauge group is non-Abelian, such
as SU(3) (for QCD), a lowest energy solution has been a problem. It is found that
all quantizations around background fields are prone to decay.
Indeed, the simplest such background is that of a uniform "magnetic" field.
The reason for studying such backgrounds for
QCD, describing quarks, the (ultimate?) constituents of all matter, is that it does not
allow the quarks to live independently which therefore cannot be seen--a
property called confinement. Their existence is only inferred.
It has been suggested that the QCD vacuum is a colour superconductor causing the
color electric fields to be expelled in analogy with the Meissner effect causing
magnetic flux to be expelled in an ordinary superconductor(leading to confinement of quarks).
To be able to understand such phenomena, it is important to study the effects of background
magnetic fields on the QCD vacuum.
The uniform background field causes the QCD vacuum to be unstable. Attempts to render
the vacuum stable have focussed on what may be called randomization of the background field.
Such a scenario is possible since most of the gauge theories of interest allow vortex solutions$[7]$,$[10]$.
The background is then to be viewed as a random configuration of vortices and the quantization is carried around this background.
However this is not an easy task and is yet to be carried out in its full glory through effective action methods.
Our approach to this problem is to utilize the fact that chaos exists in non-Abelian gauge theories
leading to the possibility of searching for an attractor in the phase space of the gauge theory.
Such an attractor may allow us to find a background configuration around which to construct a quantum theory.

Therefore it is very important to study and understand the solution spaces of
Yang-Mills theories (including topological theories such as Chern-Simons).
The purely mathematical work of Donaldson does precisely this for pure Yang-Mills
theories.

Pure Yang Mills is described by an action:
\begin{equation}
S_{YM}=-\frac{1}{2}\int_{M} Trace(F_{\mu \nu} F_{\mu \nu}).
\end{equation}
and the equations of motion are:
\begin{equation}
\nabla_{\mu} F_{\mu\nu}=0,
\end{equation}
with
\begin{equation}
F_{\mu\nu} =\partial_{\mu} A_{\nu} -\partial_{\nu} A_{\mu} +[A_{\mu},A_{\nu}].
\end{equation}
These are a set of highly non-linear PDE's for the gauge potentials
$A_{\mu}$.

Until 1974, the YM Action was thought to be the most
general action that could be written down (modulo boundary integrals) for the
YM equations in dim(M). This changed when Chern and Simons produced a
purely differential geometric result in 1974, showing the existence of
an odd-dimensional secondary characteristic class.
Coupled with the geometric interpretation of YM fields, it follows
that in odd dimensions, i.e, for odd dim(M), the YM action allows an additional
term (now called the Chern-Simons term or CS form),
modifying the Yang Mills equations.
Inclusion of this term in the action leads to extra non-linearities in the gauge
fields
\begin{equation}
S_{odd}=\frac{m}{2}\epsilon^{\mu\nu\rho}Trace(F_{\mu\nu}A_{\rho}-\frac{2}{3}iA_{\mu}A_{\nu}A_{\rho})
\end{equation}

\noindent
"Matter" fields in interaction with the gauge fields are included
giving a set of functions $(f_{1}(x), \cdots , f_{m}(x))$ on a space time manifold
transforming as an m-dimensional representation of  the gauge group G.
An action is constructed by including the $f(x)'s$. Consider , for
example, a set of scalar functions $\phi(x)$ transforming under some
irreducible representation of G. Such fields are called scalar or Higgs fields.
A typical action for the $\phi(x)'s$ in interaction with gauge fields with
gauge group G is given by:
\begin{equation}
S=\int -\frac{1}{2} Tr (F_{\mu\nu}F_{\mu\nu}) + (D_{\mu})^{\dagger}(D_{\mu}\phi) - m^{2} \phi \phi^{\dagger} +V(\phi)
\end{equation}
where
\begin{equation}
(D_{\mu})_{ij}=\delta_{ij}\partial_{\mu}-iA^{a}_{\mu}T^{a}_{ij} \cdots ; ij=1,\cdots ,m.
\end{equation}
$V(\phi)$ is a potential term which takes care of the non-linear interactions
of $\phi's$ amongst themselves.

The procedure adopted to study
Dynamical systems require the reduction of these PDE's to ODE's. This reduction
is through an assumption of spatial homogeneity
of the gauge potentials  i.e.
\begin{equation}
A_{\mu}(x)=A_{\mu}(t),
\end{equation}
The gauge potentials are assumed to be functions depending only on the time
coordinate while still taking values in the Lie Algebra L(G).
Further simplifications occur by assuming G to be the simplest
non-Abelian group : SU(2), and, through trivial dependence on the
group indices.
Writing the gauge potential as $A_{\mu}^{a}(t)$ ; a=(1,2,3) is the group index
for G=SU(2), the following is assumed:
\begin{equation}
gA_{\mu}^{1}=x(t); gA_{\mu}^{2}=y(t); gA_{\mu}^{3}=z(t),
\end{equation}
for $\mu=1,2,3$,
while
\begin{equation}
A_{0}^{a}=0 \forall a.
\end{equation}
This last condition is a simple gauge condition.
With these definitions, $H_{YM}$ becomes
\begin{equation}
H=\frac{1}{2}(p_{x}^{2}+p_{y}^{2}+p_{z}^{2})+\frac{1}{2}(x^{2}y^{2}+y^{2}z^{2}+z^{2}x^{2})
\end{equation}
and the corresponding equations of motion are:
\begin{eqnarray}
\ddot{x}&=&-x(y^2+z^2) \nonumber \\
\ddot{y}&=&-y(x^2+z^2) \nonumber \\
\ddot{x}&=&-z(y^2+x^2)
\end{eqnarray}

Thus pure Yang Mills classical dynamics corresponds to a system of coupled
oscillators via quartic potentials.
While the pure YM system is mixing, the addition of Higgs fields to the YM
system modifies the behaviour dramatically.
 The Higgs fields tend to make the
motion more regular. An additional parameter, $\lambda$, allows
altering the strength of the nonlinear coupling of the Higgs fields.

Further, when the Higgs fields are introduced, there
is a transition from regular motion to chaos, when the ratios of the Higgs
coupling to the gauge
coupling and energy of the system are varied [2].
It has been conjectured by Nikolaevskii and Shur$[8]$ that if chaos is present in
the dynamics of spatially homogeneous fields  then it is present in the full
field theory [3].
This was confirmed in the Yang-Mills field theory by considering
its classical solutions, including the Wu-Yang monopole
and reducing them to a Fermi-Pasta-Ulam problem of coupled anharmonic
oscillators [4].The analysis of the Yang-Mills Higgs system  with
't-Hooft-Polyakov monopole solutions has
also been carried out along similar lines. It has been demonstrated
that there is a transition from regularity to chaos in this system [5].
Thus the  pure Yang-Mills  and Yang-Mills Higgs theories exhibit `field theoretic'
or `spatio-temporal' chaos, thereby
 providing examples of field theories which possess two
extremes of non-linear behaviour- solitons and chaos.
 The Abelian Higgs model (which
is the relativistic extension of the Ginzburg - Landau theory) is another example
where both chaos and vortices exist.

The Maxwell Chern- Simons Higgs system (MCSH)
has  been receiving considerable attention lately as it is a potential candidate
for an effective field theory of High- $ T_{c}$ superconductors.
It has been shown that the MCSH system
(which includes the kinetic energy term for the gauge field) and the
pure Chern-Simons Higgs (CSH) system (which is the low-energy limit of MCSH) both
admit charged vortex solutions [7].
We examine here, the possibility of these systems exhibiting chaos
as in the examples cited above.

For the CSH theory we will
see below that for the case of spatially homogeneous solutions, the
field theory reduces to a dynamical system with two degrees of freedom, which is integrable. A
Painlev\'{e} analysis of MCSH
shows that the addition of the Maxwell term destroys the integrability of the system.

The Lagrangian density of the (2+1 dimensional) CSH system is given by:
\begin{equation}
L_{o}=\frac{m}{2}\epsilon^{\mu \nu \rho}A_{\mu}F_{\nu \rho} + \left | (\partial_{\mu} -ieA_{\mu}\phi)\right |^{2}-V(\phi)
\end{equation}
 By choosing the gauge $A_{o}=0$ and considering spatially homogeneous solutions $\partial_{i}\phi = \partial _{i}A_{j}=0, (i,j=1,2)$, we reduce the problem to a mechanical system.
Writing $A_{1}$=C cos($\chi$) and $A_{2}$=C sin($\chi$)
implies :
\begin{eqnarray}
\dot{\chi} & = & \frac{-e^{2}}{m}\phi^{2} \\
\ddot{\phi} + e^{2}C^{2}\phi & = & \frac{-1}{2}\frac{\partial V}{\partial \phi}
\end{eqnarray}
Therefore, the CSH theory is reduced to a dynamical system with two degrees of freedom
$ \phi $ and $ \chi $.
The corresponding Hamiltonian:
\begin{equation}
H=\frac{p_{\phi}^{2}}{4}-\frac{e^{2}}{m}p_{\chi}\phi^{2} + e^{2}C'^{2}\phi^{2} +V(\phi)
\end{equation}
$p_{\chi}$ is clearly a constant of motion. Hence, the CSH system with two
degrees of freedom, $\phi$ and $\chi$, is integrable with two integrals of motion
 H and $p_{\chi}$.

The MCSH Lagrangian is of the form:
\begin{eqnarray}
&L=-\frac{1}{4} F_{\mu\nu}F^{\mu\nu} +\frac{1}{2}m\epsilon^{\mu\nu\rho}A_{\mu}F_{\nu\rho}\nonumber \\
&+\left | (\partial_{\mu}-ieA_{\mu}\phi^{2}) \right |^{2}-V(\phi)
\end{eqnarray}
Considering the spatially homogeneous solutions:
$A_{1}=A_{1}(t)$; $A_{2}=A_{2}(t)$; $\phi=\phi(t)=|\phi|e^{i\xi}$ and choosing
the gauge $A_{0}=0$, we find once again, that $\phi$ can be chosen to be real and  that the equations of
motion can be found from the following
Hamiltonian:
\begin{eqnarray}
&H=\frac{1}{2}[(p_{1}-mA_{2})^{2} +(p_{2}+mA_{1})^{2}]+\frac{p_{3}^{2}}{4}\nonumber \\
&+e^{2}(A_{1}^{2}+A_{2}^{2})\phi^{2} +V(\phi) \nonumber
\end{eqnarray}
Using the  quartic Higgs potential $V(\phi)=\frac{\lambda}{4}(\phi^{2}-v^{2})^{2}$
and defining
$A_{1}=ACos\zeta$ and $A_{2}=ASin\zeta$ . Then the variables are A, $\phi$ and $\zeta$ and the
Hamilton's equations of motion are:
\begin{eqnarray}
\dot{A} & = & p_{A}\\
\dot{p_{A}} & = & -2e^{2}\phi^{2}A+\frac{(p_{\zeta})^{2}}{A^{3}}-m^{2}A\\
\dot{\phi} & = & \frac{p_{\phi}}{2}\\
\dot{p_{\phi}} & = & -2e^{2}A^{2}\phi-\lambda \phi(\phi^{2}-v^{2})\\
\dot{\zeta} & = & m- \frac{p_{\zeta}^{2}}{A^{3}}\\
\dot{p_{\zeta}} & = & 0.
F\end{eqnarray}
Thus we see that there are two constants of motion H and $p_{\zeta}$, but three degrees of freedom.
Now $\zeta$ is a cyclic coordinate whose dynamics is determined by the other co-ordinates and we need not
consider it further . We use the integral of motion  $p_{\zeta}$ to reduce these equations to
the second order differential equations:
\begin{eqnarray}
\ddot{\phi}=-e^{2}\phi A^{2}-\frac{\lambda}{2}\phi (\phi^{2}-v^{2})\\
\ddot{A}=-m^{2}A+\frac{p_{\zeta}^{2}}{A^{3}}-2e^{2}A^{2}\phi^{2}
\end{eqnarray}

The Lagrangian for the non-Abelian (SU(2))
 Chern Simons Higgs (NACSH) system in 2+1 dimensions
in Minkowski space is given by:
\begin{equation}
L = \frac{m}{2}\epsilon^{\mu\nu\lambda}[F^{a}_{\mu\nu}A^{a}_{\alpha} -\frac{g}{3}f_{abc}A^{a}_{\mu}A^{b}_{\nu}A^{c}_{\alpha}]+D_{\mu}\phi^{\dagger}_{a}D^{\mu}\phi_{a}-V(\phi)
\end{equation}
where:
\begin{equation}
F^{a}_{\mu\nu}= \partial_{\mu}A_{\nu}^{a}-\partial_{\nu}A_{\mu}^{a}+gf_{abc}A^{b}_{\mu}A^{c}_{\nu}
\end{equation}
$f_{abc}$ are  the structure constants of SU(2) Lie algebra
and
\begin{equation}
D_{\mu}\phi_{a}=(\partial_{\mu}-igT^{l}A^{l}_{\mu})\phi_{a}
\end{equation}
Where: $T_{a} $ are the generators of the SU(2) algebra , so that  $ tr[T_{a}T_{b}]=\lambda\delta_{ab} $
The  equations of motion can be described with three parameters and are:
parameters.
\begin{equation}
\vec{\dot{A_{1}}}=[\vec{A_{2}}\vec{\phi^{2}}-\vec{\phi}(\vec{A_{2}}\cdot\vec{\phi})]
\end{equation}
\begin{equation}
\vec{\dot{A_{2}}}=-[\vec{A_{1}}\vec{\phi^{2}}-\vec{\phi}(\vec{A_{1}}\cdot\vec{\phi})]
\end{equation}
\begin{eqnarray}
\vec{\ddot{\phi}}&=-[(\vec{A_{1}^{2}}+\vec{A_{2}^{2}})\vec{\phi}\nonumber \\
&-(\vec{A_{1}}\cdot\vec{\phi}\vec{A_{1}}+\vec{A_{2}}\cdot\vec{\phi}\vec{A_{2}})]\nonumber \\
&-\frac{\kappa}{2}\vec {\phi}(\vec{\phi}^{2}-v^{2})
\end{eqnarray}

with $\kappa=\frac{\lambda m}{g^{2}}$ .
We shall set the scaled $v$ to be one without loss of generality.

The Yang-Mills Chern-Simons Higgs System(YMCSH)  Lagrangian is:
 \begin{eqnarray}
L&=&-\frac{1}{4}F_{\mu\nu}^{a}F^{\mu\nu a}+\frac{m}{2}\epsilon^{\mu\nu\alpha}[F_{\mu\nu}^{a}A^{a}_{\alpha}-\frac{g}{3}f_{abc}A^{a}_{\mu}A_{\nu}^{b}A^{c}_{\alpha}]\nonumber \\
&&+D_{\mu}\phi^{\dagger}_{a}D^{\mu}\phi_{a}-V(\phi)
\end{eqnarray}
The YMCSH dynamical equations
in the gauge $A_{0}=0$ andin the spatially homogeneous
 case are:
\begin{eqnarray}
\frac{1}{m}\ddot{\vec{A_{1}}}+2\dot{\vec{A_{2}}}+2(\vec{A_{1}}\vec{\phi^{2}}
-\vec{\phi}\vec{A_{1}}\cdot \vec{\phi})\nonumber \\
+\frac{1}{m}(\vec{A_{1}}\vec{A_{2}}\cdot\vec{A_{2}}-\vec{A_{2}}\vec{A_{1}}
\cdot\vec{A_{2}})=0\\
\frac{1}{m}\ddot{\vec{A_{2}}}-2\dot{\vec{A_{1}}}+2(\vec{A_{2}}\vec{\phi^{2}}
-\vec{\phi}\vec{A_{2}}\cdot \vec{\phi})\nonumber \\
+\frac{1}{m}(\vec{A_{2}}\vec{A_{1}}\cdot\vec{A_{1}}-\vec{A_{1}}\vec{A_{1}}
\cdot\vec{A_{2}})=0
\end{eqnarray}
and
\begin{eqnarray}
\ddot{\vec{\phi}}&=-[(\vec{A_{1}^{2}}+\vec{A_{2}^{2}})\vec{\phi}-(\vec{A_{1}}\nonumber \\
&\cdot\vec{\phi}\vec{A_{1}}+\vec{A_{2}}\cdot\vec{\phi}\vec{A_{2}}] \nonumber \\
&-\frac{\kappa}{2}\vec{\phi}(\vec{\phi}^{2 }-1).
\end{eqnarray}
It is interesting to note that while in the NACSH system the Yang-Mills
parameter
 g, the Higgs parameter $\lambda$ and the Chern-Simons parameter m could all be
 combined
into the parameter $\kappa$, this is not possible for the YMCSH system where we
 are left with both $\kappa$ and m appearing explicitly.
The  energy function is:
\begin{eqnarray}
&E=\frac{1}{2m}(\dot{\vec{A_{1}^{2}}}+\dot{\vec{A_{2}^{2}}})\nonumber \\
&+\vec{\dot{\phi}}^{2}+\frac{1}{2m}[\vec{A_{1}^{2}}\vec{A_{2}^{2}}-(\vec{A_{1}}\cdot\vec{A_{2}})^{2}]\nonumber \\
&+[(\vec{A_{1}^{2}}+\vec{A_{2}^{2}})\vec{\phi^{2}}-(\vec{A_{1}}\cdot\vec{\phi})^{2}\nonumber \\
&-(\vec{A_{2}}\cdot\vec{\phi})^{2}] +\frac{\kappa}{4}(\vec{\phi}^{2}-1)^{2}.
\end{eqnarray}
This completes the description of the dynamical systems which we shall be
studying.

\subsection{Painlev\'{e} Analysis.}
The Painlev\'{e} analysis
for the CSH system  is instructive in illustrating the Painlev\'{e} property and serves
as a useful precursor to test the integrability property of the more complicated
MCSH theory.

The Painlev\'{e}  test for integrability is usually stated as follows$[11]$,$[12]$:
Consider the system of differential equations:
\begin{equation}
 \frac{d^{n_{i}}x_{i}}{dt^{n_{i}}}=f_{i}[t;x_{i},\dot x_{i} , x_{2},\dot x_{2}, \cdots]\\
\end{equation}
and continue $x_{i}(t)$ from real to complex times. Then, the Painlev\'{e}
conjecture asserts that if the singularities of $x_{i}(t)$ are no more than poles or branch points  and if the
Laurent expansion around the leading singularity has a sufficient number of arbitrary constants
warranted by the set of differential equations, then the system is integrable.
The actual Painlev\'{e} analysis proceeds in three steps:\\
1) Determine the leading singularity.
If the leading singularity is a pole ( branch point) it indicates a strong (weak) Painlev\'{e} property.
If neither is the case, then the system is non-integrable.\\
2. If the leading singularity is a pole or a branch point, a formal Laurent series expansion of the solution around the singularity $t_{0}$ is carried out.
The powers of $(t-t_{0})$ in the series expansion,  for which the corresponding coefficients become arbitrary,  are  determined.
 These are called the Kowaleskaya exponents or resonances .\\
3.) Next we verify that at the resonance values sufficient number of arbitrary constants exist.

For the CSH system the equation of motion can be written as:
\begin{equation}
\ddot{x}=-x-{\cal A}x(x^{2}-1).
\end{equation}
To find the dominant term we continue t to the complex plane and substitute:
$x=a(t-t_{0})^{-\alpha}.$
We find that $\alpha=1$ and $a^{2}=2-2/{\cal A}$.
Thus the leading singularity is a pole.

Including the next to leading term :
$ x= a(t-t_{0})^{-1} + p(t-t_{0})^{r-1} $
and balancing the terms linear in p, we find that the roots are r=-1 and r=4.
Carrying out the Laurent series expansion about the leading singularity,
\begin{equation}
x=\sum_{i=0}^{\infty}a_{i}(t-t_{0})^{i}
\end{equation}
we find that all the $a_{i}$'s are determined except $a_{4}$, which is arbitrary.
Thus the second order differential equation has a solution with two arbitrary constants $t_{0}$ and $a_{4}$ .
Hence, the system is integrable.

We can now check the integrability (or lack of it) of the MCSH system by carrying out the Painlev\'{e} test .
The equations of motion we examine are the second order differential equations written above.
In terms of the rescaled variables $x=\frac{\phi}{v}$, $y=\frac{A}{v}$, t'=evt, ${\cal C}=\frac{p_{\zeta}^{2}}{e^{2}v^{6}}$
${\cal A}=\frac{\lambda}{2e^{2}}$, ${\cal B}=\frac{m^{2}}{e^{2}v^{2}}$, we get the
equations:
\begin{eqnarray}
\ddot{x}=-xy^{2}-{\cal A}x(x^{2}-1)\\
\ddot{y}=-{\cal B}y +\frac{{\cal C}}{y^{3}}-2x^{2}y
\end{eqnarray}
where the differentiation is w.r.t t'. In the subsequent analysis we drop the prime.
The dominant singularity structure is found by substituting:\\
$x=a(t-t_{0})^{\alpha}$ and $y=b(t-t_{0})^{\beta}$,\\
where $\alpha$ and $\beta$ are assumed to be less than zero.
Balancing the singularity at $t_{0}$ gives:
\begin{eqnarray}
&a\alpha(\alpha-1)(t-t_{0})^{\alpha-2}=-ab^{2}(t-t_{0})^{\alpha+2\beta}\nonumber \\
&- {\cal A}a^{3}(t-t_{0})^{3\alpha}+{\cal A}a(t-t_{0})^{\alpha}\nonumber \\
&b\beta(\beta-1)(t-t_{0})^{\beta-2}=-{\cal B}b(t-t_{0})^{\beta}\nonumber \\
&-2a^{2}b(t-t_{0})^{\alpha+2\beta}\nonumber
\end{eqnarray}
These equations  immediately reveal that  there are two possibilities for the leading order :
\begin{itemize}
\item Case 1. $\alpha=-1$, $\beta=-1$\\
$a^{2}=-1$ and $ b^{2}={\cal A}-2 $
\item Case 2. $\alpha=-1$ and $\beta=\frac{1}{2} \pm \frac{1}{2}\sqrt{1+16/{\cal A}}$
 ; $a^{2}=-\frac{2}{{\cal A}}$ and b arbitrary.
\end{itemize}
 Both cases must be tested for the Painlev\'{e} property.

The resonance analysis is carried out for both cases below.\\
Case 1:
\begin{eqnarray}
x=a(t-t_{0})^{-1}+p(t-t_{0})^{r-1}\\
y=b(t-t_{0})^{-1}+q(t-t_{0})^{r-1}
\end{eqnarray}
with $a^{2}=-1$ and $b^{2}={\cal A}-2$
where p and q are arbitrary parameters.
From the  above equations , we find that the resonances
occur at\hspace{1cm}r=-1,4,$\frac{3}{2}\pm \frac{\sqrt{8{\cal A}-7}}{2}$.\\
Reality of the roots  together with the non- leading nature of the resonance terms gives the condition $\frac{7}{8}\le {\cal A} \le 2$.\\
For this case to have the Painlev\'{e} property we require r to be a non-negative integer (except for r=-1, which is the root cooresponding
to the movable singularity).
This gives rise to two possibilities:
\begin{itemize}
\item case 1(a).
\begin{equation}
\sqrt{8{\cal A}-7}=1\hspace{1cm};{\cal A}=1\hspace{1cm};r=-1,1,2,4.
\end{equation}
\item case 1(b).
\begin{equation}
\sqrt{8{\cal A}-7}=3\hspace{1cm};{\cal A}=2\hspace{1cm};r=-1,4,3,0.
\end{equation}
\end{itemize}
We check the Painlev\'{e} property of cases 1(a) and 1(b) by substituting
\begin{equation}
x=\sum_{j=0}^{\infty}a_{j}(t-t_{0})^{j-1}
\end{equation}
and
\begin{equation}
y=\sum_{j=0}^{\infty}b_{j}(t-t_{0})^{j-1}
\end{equation}
For case 1(a), our resonance analysis indicates that the indeterminate coefficients
in these Laurent expansions should occur at j=1,2 and 4. If $a_{1}/b_{1}$ , $a_{2}/b_{2}$,
$a_{4}/b_{4}$ are arbitrary, then the system is integrable.
When we carry out the detailed analysis for case 1(a), we find:\\
$b_{1}^{2}=4a_{1}^{2}$,
and
$a_{1}^{2}=({\cal {\cal B}}+1)/9$\\
Thus $a_{1}$ and $b_{1}$ are determined and the Painlev\'{e} property fails.
Furthermore , for $j\le 4$ all the $a_{j}$'s and $b_{j}$'s are determined
 except $a_{4}$ and $b_{4}$, between which there is a linear relation.
 For $j\ge 4$ , the coefficients can be determined from the $a_{j},b_{j}......j=1,..4$ . Thus there is only one arbitrary constant
among the $a_{j}$'s and the $b_{j}$'s.But in order for case 1(a) to have the Painlev\'{e}
property there must be at least 3 arbitrary parameters beside $t_{0}$. Thus,case 1(a) fails the Painlev\'{e} test.

For case 1(b), r=0 is a root. This implies that $a_{0}$ and $b_{0}$ should be arbitrary.
However, in this case, we find that $a_{0}^{2}=-1$ and $b_{0}^{2}={\cal A }-2=0$, 1.e they are determined. This , therefore,leads to a contradiction
and the system does not pass the Painlev\'{e} test.

Case 2:\\
In this we find that the resonances occur at r=-1,0.4, $1-2\beta$
In order that $r>0$ ,we have the condition that $1-2\beta>0$ . As $\beta >-1$
and $1-2\beta=m$ ,
 $m=0,1,2$.\\
Thus,we have three cases.
\begin{itemize}
\item Case 2a.\\
m=0\hspace{1cm};$\beta=\frac{1}{2}$; \hspace{1cm} $1+\frac{16}{{\cal A}}=0$ or ${\cal A}=-\frac{1}{16}$. \\This value of${\cal A}$ is unphysical as it corresponds to
an unbounded Higgs potential.
\item Case 2b.\\
m=1;\hspace{1cm}$\beta=0$; \hspace{1cm} ${\cal A}=\infty$.\\
 Again, this is not physically interesting as for ${\cal A}=\infty$ the Higgs potential is infinite.
\item Case 2c.\\
m=2;\hspace{1cm} $\beta=-1/2$. Hence,  we examine the system for the `weak Painlev\'{e} property'.\\
 In this case,  ${\cal A}=\frac{16}{3}$ and r=-1,0,2,4.\\
We now carry out the consistency check of the full resonance analysis by substituting:
\begin{equation}
x=\sum_{0}^{\infty}a_{j}(t-t_{0})^{j-1}
\end{equation}
\begin{equation}
y=\sum_{0}^{\infty}b_{j}(t-t_{0})^{j-\frac{1}{2}}
\end{equation}
 We find that $a_{0}^{2}=\frac{-2}{{\cal A}}$ and that $a_{1}$ and $b_{1}$ are determined in terms of $b_{0}$.In the second order  $b_{0}$ is determined in terms of ${\cal B}$.
 But , as r=0 is a root , one of the two coefficients $a_{0}$ or $b_{0}$ must be arbitrary. Hence,case 2c also fails the
Painlev\'{e} test .
\end{itemize}
We have thus established that for all possible cases the MCSH theory is non-integrable.

\section{Order -Chaos Transitions}
 Is there a sharp order to chaos transition in the parameter space of these theories?
 In the context of Abelian
Higgs theories Kawabe $[7]$ has reported transition from order to chaos
within a certain range of the Higgs coupling constant and energy .
The onset of chaos is remarkably different qualitatively from the corresponding
transition in the YMCS system where
 the existence of an interesting `fractal' structure appears in the phase
transition region.
This aspect of chaos in YM systems studied in  the non-Abelian CSH (NACSH) and the
YMCSH systems with an SU(2) symmetry group. A comparative analysis is done to see
the role of the kinetic term, the Higgs term and the CS term in the transition.

We have computed the Lyapunov exponents for various values of the parameters $\lambda$, $e^{2}$, m and v.  In each case  the
Lyapunov exponent $[12]$ converges to some positive value.
We examined the variation of the maximal Lyapunov exponent as the two
NACSH parameters
energy and $\kappa$ are varied. This clearly shows us regions of regular
behaviour (where the exponent goes to zero)
and regions of chaotic behaviour (where the exponent is positive). These
calculations were carried out for a wide range of initial
conditions. The initial conditions that were chosen were in turn dictated by the
 dynamical systems themselves.
  Being derived from the
equations of motion the field variables are required to satisfy the Gauss' law
constraint.
Fig. 1 shows the comparision of maximal Lyapunov exponent versus x in a two-variable initial
ansatz for YMCSH for different values of $\kappa$, energy and m.
The graph shows
that for large $\kappa$ (where either the Higgs coupling $\lambda$ is large or
the YM coupling g is small) the system exhibits more regularity for low energies.
Here, we see that for the YMCSH system such transitions to regularity are seen
for small energies as $\kappa$ increases while the Lyapunov exponent increases
almost linearly with energy for the large energy regime.

A striking feature that emerges in our studies that is
counter-intuitive is that, in general, it is not true that
increase in $\kappa$ `regularizes' the gauge term at all energies.
An increase in $\kappa$ could either be due to an increase in the Higgs coupling
$\lambda$ or a decrease in the gauge coupling $g$, for a fixed m.
Whereas in the former case the regularity is expected to increase , in the
latter case, it is not established that for all small
non-zero g more regular islands appear.
{\bf An increase in $\kappa$ produces more regularity only for small
values of energy}.
This can be understood better if we realize that it is not just the value of
$\kappa$ that determines the appearance of regular islands, but
 also the available phase space as well.

Another aspect is that as the energy increases, the maximal Lyapunov exponent
increases in magnitude in the YMCSH system. This shows that the YM terms takes
over for large energies and the CS term produces the `oscillatory effect'. The
effect of the CS term is reminiscent of the `fractal' structure observed in
YMCS systems where in various energy windows, order-chaos-order transitions are observed.

The final picture which emerges bears out the fact that in a complex
dynamical system with a large phase space (in contrast to the wide class
of Hamiltonian systems with two degrees of freedom) curious interplay between
different coupling constants and the rich structure of phase space itself can
lead to novel results- some of them quite counter-intuitive and surprising.

A step towards field theoretic chaos in these systems has been the subject of our current studies.
We consider the dynamics of the  NACSH and YMCSH system  by perturbing around initial vortex solutions.
Initial studies have shown that such an exercize holds promose in yielding results of the Fermi-Pasta Ulam type for coupled
harmonic oscillators. However, we have to note that the $\frac{1}{r}$ singularity of the vortex solution has been a cause for concern in the dynamical evolution
of the solutions vis-a-vis the energy conservation.
We hope to solve this particular aspect of the problem using more refined PDE solution tools.
Another direction for further research is the study of quantum chaos in these systems and the distribution of energy level spcaing in the quantum mechanical system.

\section{REFERENCES}
1) B. Bambah,  S. Lakshmibala, M.S Sriram and C. Mukku
 Phys. Rev {\bf D47},(1993).\\
2)     C.  Mukku, B. Bambah, S. Lakshmibala, M.S. Sriram and J. Segar.
      J.Phys. {\bf A 30},  (1997) 3003 .        \\
3) S. LakshmiBala, B. Bambah , C. Mukku and M.S. Sriram .
      Pramana Journal of Physics, {\bf Vol. 48} no.2, pp 617-635,
      Special issue on " Nonlinearity and Chaos in Physics".(1997).\\
4) "Chaos in Chern-Simons Higgs Systems",
      Abstracts of the X HEP symposium held at Bombay, 1992 p. 162.\\
5) M.S Sriram,  C. Mukku, S.Lakshmibala and B. Bambah
      Phys. Rev {\bf D49} (1993) 4246-4257.\\
6) S.G. Matinyan, G.K. Savvidy and N.G.T. Savvidy, Sov. Phys. JETP {\bf 53},421,(1981).\\
   S.G. Matinyan, G.K. Savvidy and N.G.T. Savvidy, JETP lett. {\bf 34},590,(1981).\\
7) C.N.Kumar and A.Khare, J.Phys.A {\bf 22}, L849 (1989).          \\
   B.Dey, C.N. Kumar and A. Sen, Institute for Plasma Research (Ahmedabad, India) Preprint RR-48/91, (1991).\\
   T. Kawabe, Phys. Lett.B {\bf 274}, 399 (1992).\\
8) E.S.Nikolaevskii and L.N.Shur, JETP lett. {\bf 36},218,(1982). \\
9) S.G. Matinyan,E. B. Prokhorenko and G.K. Savvidy , JETP lett. {\bf 44},139,(1986).\\
10) T. Kawabe and S. Ohta , Phys. Rev. {\bf D44} , 1274 (1991).\\
   M.P.Joy and M. Sabiir, Pramana {\bf 38}, L91, (1992).\\
11) M.J. Ablowitz, A. Ramani and H. Segur, J.Math. Phys.,{\bf 21} 715 (1980).  \\
12) M.Tabor, {\bf `Chaos and Integrability in Non-Linear Dynamics '}, (John Wiley and Sons, New York,1989).
\end{document}